\documentclass[twocolumn,showpacs,preprintnumbers,superscriptaddress,amsmath]{revtex4}
\usepackage[dvips]{graphics}
\usepackage{epsfig}
\usepackage{psfrag}
\usepackage[psamsfonts]{amssymb}
\usepackage{amsmath}
\usepackage{indentfirst}
\usepackage{amssymb}
\usepackage{wrapfig}
\usepackage{url}
\usepackage{dcolumn}
\usepackage{bm}

\newcommand{\be}{\begin{equation}}
\newcommand{\ee}{\end{equation}}
\newcommand{\ba}{\begin{eqnarray}}
\newcommand{\ea}{\end{eqnarray}}

\begin{document}

\preprint{APS preprint}

\title{Vere-Jones' Self-Similar Branching Model}

\author{A. Saichev}
\affiliation{Mathematical Department, Nizhny Novgorod
State University, Gagarin prosp. 23, Nizhny Novgorod,
603950, Russia} \affiliation{Institute of Geophysics and
Planetary Physics, University of California, Los Angeles,
CA 90095}

\author{D. Sornette}
\affiliation{Institute of Geophysics and Planetary
Physics and Department of Earth and Space Sciences,
University of California, Los Angeles, CA 90095}
\affiliation{Laboratoire de Physique de la Mati\`ere
Condens\'ee, CNRS UMR 6622 and Universit\'e de
Nice-Sophia Antipolis, 06108 Nice Cedex 2, France}
\email{sornette@moho.ess.ucla.edu}

\date{\today}

\begin{abstract}
Motivated by its potential application to earthquake statistics, we
study the exactly self-similar branching process introduced recently by
Vere-Jones, which extends the ETAS class of 
conditional branching point-processes of triggered seismicity. 
One of the main ingredient of Vere-Jones' model is that the 
power law distribution of magnitudes $m'$ of daughters of first-generation of 
a mother of magnitude $m$ has two branches $m'<m$ with exponent $\beta-d$
and $m'>m$ with exponent $\beta+d$, where $\beta$ and $d$ are two positive parameters.
We investigate the condition and nature
of the sub-critical, critical and super-critical regime in this
and an extended version interpolating smoothly between several models.
We predict that the distribution of magnitudes of events triggered
by a mother of magnitude $m$ over all generations has also two branches
$m'<m$ with exponent $\beta-h$
and $m'>m$ with exponent $\beta+h$, with $h= d \sqrt{1-s}$, 
where $s$ is the fraction of triggered events. This corresponds to 
a renormalization of the exponent $d$ into $h$ by the hierarchy
of successive generations of triggered events.
The empirical absence
of such two-branched distributions implies, if this model is seriously
considered, that the earth is close to criticality $s \simeq 1$ so that
$\beta - h \approx \beta + h \approx \beta$. We also find that, 
for a significant part of the parameter space, the 
distribution of magnitudes over a full catalog summed over an average
steady flow of spontaneous sources (immigrants) reproduces the distribution
of the spontaneous sources and is
blind to the exponents $\beta, d$ of the distribution
of triggered events. In summary, the exactly self-similar Vere-Jones model
provides an attractive new approach to model triggered seismicity, 
which alleviates delicate questions on the role of magnitude cut-offs in 
other non-self-similar models and provides
interesting predictions which could be tested
with stochastic reconstruction methods of earthquake sequences.

\end{abstract}


\maketitle

\section{Introduction}

Stochastic branching processes describe well a multitude of phenomena
\cite{Athreya,Sankaranarayanan} from chain reactions in nuclear and
particle physics, material rupture, fragmentation and earthquake
processes, to population and biological dynamics, epidemics, economic
and social cascades and so on. Branching processes are also of
particular interest because deep connections have been established with
critical phenomena. In branching processes, various quantities exhibit
power law distributions at criticality. This includes the distributions
of cluster sizes, of the number of generations before extinction and of
durations. 

Branching processes become critical when the average death rate
is exactly compensated by the average growth rate. At criticality,
branching processes become asymptotically self-similar,
which translates for instance into an asymptotic power law tail
for the distribution of cluster sizes. 
Such scale-invariance is a general characteristic of systems
at the critical point of a phase transition or of a bifurcation.
But self-similarity in general holds only (i) at criticality and (ii)
asymptotically, i.e., at scales much larger than the 
microscopic mesh or elementary branch scale.

Vere-Jones has recently introduced a class of self-similar branching
processes which is exactly self-similar for a broad range of parameters
(of non-zero measure), that is, far from criticality and for all scales
\cite{VJmodel}. Vere-Jones self-similar branching process is
derived from a class of exactly self-similar random measures which
generalize the class of stable purely atomic completely random
measures. The underlying idea is that a change of scale is balanced by a
change in mass, making the measure, and the branching process,
self-similar at all scales and off-criticality. This is 
possible only for branching processes with continuous masses or ``marks.''
Then, the model has a natural application to describe 
earthquake triggering, which is the example we adopt in the following
to formulate the problem and present our results,
without loss of generality.

The concrete example proposed by Vere-Jones is the ``self-similar ETAS''
(epidemic-type aftershock sequence) model, which is a self-similar
extension of the initial non-self-similar standard ETAS model
\cite{Ogata,KK81}. The main statistical properties of the standard ETAS
model are reviewed in \cite{HS02,Saichevetal04}. The ETAS model describes the rate
$\lambda({\vec r},t,m)$ of events (earthquakes for instance) at
position ${\vec r}$ at time $t$ and of mass or mark (magnitude) $m$
resulting from spontaneous sources (``immigrants'') and from all past
events. The model is defined in terms of the conditional Poisson
intensity $\lambda({\vec r},t,m)$ which is a function of all past
events. The standard ETAS model describes earthquake triggering by past
ones within the framework of branching theory and takes into account the
interplay between the exponential productivity law and the
Gutenberg-Richter (GR) law of event sizes. A series of
papers has shown that the standard ETAS model
gives a reasonable description of the statistics of seismic clustering
\cite{Ogata}, of foreshocks \cite{HSGth,Forexp} and
aftershocks \cite{HS02,Saichevetal04} triggered by other earthquakes,
of the empirical B{\aa}th's law  for the largest aftershock
of a given sequence \cite{bathnum,SSBaththe} and of the
statistics of seismic rates \cite{SaiSorpdf}. The
standard ETAS model thus constitutes a powerful null hypothesis
to test against other models \cite{Ogata}.

The standard ETAS model is perhaps the simplest of a much larger class
of models embodying the physics of triggered seismicity. Its simplicity
results from its assumption of complete decoupling between the
Gutenberg-Richter distribution of event sizes, productivity law, time
and space interaction kernels. Already in 1988, Ogata proposed an
extension allowing for magnitude-dependent time and space kernels and
taking into account spatial anisotropic effects \cite{Ogata}. Using statistical
likelihood methods, this extension has recently been
shown to be superior to the standard ETAS model to account for
empirical earthquake clustering \cite{OgataZhuang}.
Ogata et al. have also extended ETAS to characterize regional features of
seismic activity in and around Japan, by allowing the parameter values
to vary from place to place \cite{Ogatakattan}. In the same vein,
we have shown that the empirical distribution of seismic rate in California
can be described adequately by ETAS model only when taking into account
a strongly non-uniform fractal distribution of spontaneous sources
\cite{SaiSorpdf}. Ouillon and Sornette have considered a multifractal
model of triggered seismicity which predicts magnitude-dependent
exponents for the Omori law of aftershock decay rates, in good
agreement with empirical data \cite{SO05_1,OS05_2}. Unfortunately,
few of these extensions have allowed yet for a full theoretical understanding
of the stationary properties of the resulting catalogs of triggered events.

Here, we present a detailed theoretical understanding of some
statistical properties of the self-similar Vere-Jones model, which
extends the standard ETAS model by allowing for some dependence between
the random magnitudes of progenies and ancestors. In an effort to
clearly formulate some universal key features of the self-similar
Vere-Jones model, we shall consider mostly the statistics of total
numbers of aftershocks triggered by spontaneous sources and their
distribution. 

The organization of the paper is as follows. Section 2 defines
Vere-Jones' model together with a generalized version that allows
us to interpolate smoothly between the standard ETAS model and Vere-Jones' model.
Section 3 presents the general theoretical treatment in terms of
generating probability functions (GPF) which predicts in particular the exact
form of the magnitude distributions. Section 4 explores the phase
diagram of the model by identifying the conditions for sub-criticality,
criticality and super-criticality. Section 5 derives the distributions of
earthquake magnitudes for Vere-Jone's self-similar model. Section 6 presents
a summary of our predictions,
a discussion of their consequences and their possible empirical tests.

\section{Vere-Jones self-similar ETAS model}

\subsection{Definition of the  ``generalized Vere-Jones ETAS model'' \label{genevj}}

Here, we integrate
over space and time to focus on global properties, such as the total number
of events triggered by a given event. Let us consider first a model
more general than the Vere-Jones ``self-similar ETAS'' model,
which we call the ``generalized Vere-Jones ETAS model.'' 
In this model, an event of magnitude $m$ may trigger another event
of magnitude $m'$ according to the Poisson rate $\lambda(m,m')$,
so that the total rate of production of events of magnitude $m'$ is
\be
\lambda(m') = \sum_{m} \lambda(m,m')
\ee
where
\be
\lambda(m,m')=\kappa \, e^{a m- \beta m' -d |m-m'|} ~,
\label{13} 
\ee
gives the number of events of magnitude $m'$ triggered directly
(on first generation) by an event of magnitude $m$. The space
and time dependence of the mother (or triggering event) of magnitude $m$
and daughter (or triggered event) of magnitude $m'$ have disappeared
from the expression of $\lambda(m,m')$ due to the integration over
space and time, so that we are concerned only here with total numbers
in a fixed space-time window. The exponential in (\ref{13}) contains
three contributions: 
\begin{enumerate}
\item  $e^{a m}$ describes the exponentially growing productivity of a
source as a function of its magnitude $m$ \cite{HelmPRL,HelmKaganJackson04}
(in other words, $e^{a m}$ is
proportional to the average number of first-generation progenies  of the
source $m$). 
\item $e^{-\beta m'}$ is the so-called Gutenberg-Richter
distribution of the magnitudes of the first-generation triggered events,
usually improperly referred to in the seismological literature as
``aftershocks'' and which is now progressively replaced by the concept
of triggered earthquakes. With this caveat in mind, we shall sometimes
use the term ``aftershocks'' in a broader sense than used in the
standard literature to refer to triggered events of arbitrary magnitudes
without the condition that the aftershock magnitudes have to be smaller
that the source magnitude.
\item The new term $e^{-d|m-m'|}$, compared with 
previous models of the ETAS class, describes a binding or localization of the
magnitude $m'$ of triggered events in the neighborhood of the ancestor's
magnitude $m$. In other words, this term means that 
daughters' magnitudes keep a memory of the size of their mothers: 
large (small) daughters come more probably from large (small) mothers for $d>0$.
This could be associated with the fact that triggered events occur
on patches of the mother's fault rupture with large residual stresses (for the 
$e^{-d(m-m')}$ branch with $m'<m$) and on faults branching from the mother's rupture
(for the branch $e^{d(m-m')}$ with $m'>m$).
\end{enumerate}
The branching model is such that any event can trigger other events 
according to the rate $\lambda(m,m')$ given by (\ref{13}). Thus, a given
event may give daughters of first-generation, which can themselves trigger
other events and so on, giving rise to a triggering cascade.
The model also considers the existence of spontaneous sources (immigrants),
seeding the branching cascades.

\subsection{The ETAS model: $d=0$ \label{etasstan}}

For $d=0$, i.e., when the size of daughters is independent (has no memory)
of the size of the mother, model (\ref{13}) recovers the standard ETAS
model \cite{Ogata,KK81,HS02}:
\be
\lambda(m,m')= \kappa' e^{a (m-m_0)} \times \beta e^{-\beta (m'-m_0)}~,
\label{nhgjle}
\ee
where the first factor in the r.h.s is the so-called exponential
productivity law and the second one is the Gutenberg-Richter law of
first generation events magnitudes which is normalized
($\int_{m_0}^{\infty} dm' ~\beta e^{-\beta (m'-m_0)}=1$). A minimum event size $m_0$ is 
necessary to make the ETAS model well-defined \cite{SorWer1,SorWer2}.
Indeed, it is not possible to make the model convergent for $m_0 \to -\infty$
(a $-\infty$ magnitude corresponds to a vanishing energy, since
the magnitude is proportional to the logarithm of the energy), and
an ultra-violet cut-off is necessary \cite{SorWer1,SorWer2}.

The constant factor $\kappa'$ is derived from the coefficient 
$\kappa$ of the generalized Vere-Jones ETAS model (\ref{13}) as
\be
\kappa'= {\kappa \over \beta}~ e^{(a-\beta) m_0} ~.
\ee
The ETAS model is critical when its average branching ratio
(equal to the number of progenies averaged over all mothers' magnitudes)
\be
n \equiv \int_{m_0}^{\infty} dm \kappa' e^{a (m-m_0)} \times \beta e^{-\beta (m-m_0)}
= {\kappa' \beta \over \beta - a}
\label{mgmlwlws}
\ee
is unity. The case $n<1$ (respectively $n>1$) corresponds to the sub-critical
(respectively super-critical) regime. The condition $a < \beta$
is needed to make the model convergent, as both the borderline $a=\beta$
and the regime $a > \beta$ leads to finite-time singularities with 
stochastic times \cite{SHfts}. Alternatively, convergence and stationary
is obtained by adding an upper magnitude cut-off $m_{\rm max}$, 
associated with the empirical bending down of the Gutenberg-Richter law
for magnitudes larger than about $8$, suggesting that $m_{\rm max} \approx 8-9$
\cite{kagantaper,Pissor1,Pissor2}.
We shall not consider the influence of this upper cut-off whose impact
is rarely felt only at time scales and for earthquake numbers so large to
ensure that the largest earthquakes are sampled.

The ETAS model, unlike Vere-Jones model, lacks self-similarity due to
both the non-generic condition $n=1$ for criticality and the existence
of a minimum event magnitude $m_0$ which introduces a characteristic
magnitude scale.

\subsection{The self-similar Vere-Jones model: $d>0$ and $a=\beta$ \label{jglld}}

Vere-Jones self-similar ETAS model corresponds to $a=\beta$ and $d>0$ in
(\ref{13}):
\be
\lambda(m,m')= \lambda(m-m')~, \qquad \lambda(m)= \kappa e^{\beta m- d|m|}~ . 
\label{14}
\ee
The condition $a=\beta$ expresses a balance between the
exponential small probability of finding a large mother and its
exponentially large productivity. When $a=\beta$ (and in 
absence of the effect of $d$), each event magnitude
range $[m,m+dm]$ contributes equally to other event triggering, and, in
particular, small events are as important to event triggering
as are larger ones, a property
which seems to be approximately true for earthquakes
\cite{HelmPRL,HelmKaganJackson04}. In the standard
ETAS model, the case $a=\beta$ is not possible without the introduction of both
an ultra-violet cut-off $m_0$ and an additional infra-red cut-off $m_{\rm max}$
truncating the Gutenberg-Richter distribution \cite{kagantaper,Pissor1,Pissor2}, 
which are necessary in order to obtain
non-diverging sequences. In Vere-Jones self-similar ETAS model,
the condition $a=\beta$ is made possible without cut-offs by the introduction of the
parameter $d>0$. Physically, expression (\ref{14}) (together
with the standard Gutenberg-Richter law) can be interpreted
by the existence of two branches for the Gutenberg-Richter distribution
of the $\sim e^{\beta m}$ events triggered by a given mother of magnitude $m$:
\begin{enumerate}
\item daughters with magnitude $m'<m$ have their magnitudes $m'$
distributed according to $\sim e^{-(\beta - d) m'}$ while
\item   daughters with magnitude $m'>m$ have their magnitudes $m'$
distributed according to $\sim e^{-(\beta + d) m'}$. 
\end{enumerate}
The former (respectively latter) distribution branch of daughter magnitudes
ensures that there are less small (respectively large) daughters than with $d=0$.
This is the origin, as we will make clear below quantitatively, 
of the possibility to avoid any cut-off $m_0$ or $m_{\rm max}$
and still obtain a convergent model. This is a first reason
why Vere-Jones' model (\ref{14}) is self-similar. As we shall see below,
the other reason is that there is a finite range of parameters
for which Vere-Jones is effectively critical.

\section{General theoretical formulation}

\subsection{From Bernoulli to Poisson Statistics}

One of the main ingredients of both the standard ETAS and
Vere-Jones' models of aftershocks triggering is the
Poissonian statistics governing the number of first generation
aftershocks triggered by some spontaneous source of given
magnitude $m$. In order to obtain a deeper insight into
the origin of the underlying Poissonian law, we start with 
the more general Bernoulli approach to the
description of aftershocks triggering statistics. 

Consider the Bernoulli version of a
generalized model in which each spontaneous source of magnitude
$m$ has $p$ independent ``possibilities'' to
trigger some aftershock. The probability that an aftershock
of magnitude in the interval $[m',m'+dm']$
is actually triggered along one of these $p$ paths is denoted
$D(m,m',p)dm'$.  Keeping the approach general, we allow $D(m,m',p)$
to depend on the source magnitude $m$, the number of channels $p$
and the triggered magnitude $m'$ in an arbitrary way, in all
what follows in this section. The next section will then apply
our formalism to the specifications (\ref{13}) and (\ref{14}),
using the standard ETAS version (\ref{nhgjle}) as a reference and
point of comparison.

The generating probability function (GPF) of the random number of first generation
aftershocks, obeying to the above
Bernoulli statistics, is given by
\be
\Theta_1(z;m,p)=\left[1+ \int_{m_0}^\infty dm'
D(m,m',p)(z-1)\right]^p ~ .
\label{mhml}
\ee
Here, $m_0$ is the smallest magnitude of possible
triggered aftershocks, which will be pushed to $-\infty$ 
in the self-similar Vere-Jones version.

An essential assumption of the models studied here is that
triggered events of first generation act themselves as sources
which trigger their own aftershocks according to the same laws.
Let us call $\Theta(z;m',p)$ the GPF of the random number of events
of all generations triggered by a first-generation daughter with
magnitude $m'$. Then, the GPF of the number of aftershocks 
triggered over all generations by a given mainshock of magnitude $m$ 
is solution of
\be
\Theta(z;m,p)= \\ \displaystyle \left[1+
\int_{m_0}^\infty dm' D(m,m',p)(z \Theta(z;m',p)-1)\right]^p ~,
\label{nejjs}
\ee
obtained from (\ref{mhml}) by replacing $z$ by $z\Theta(z;m',p)$
to express that each branch has the same statistical cascade properties.

Consider the limiting intensity of aftershocks triggering
\be
\lim_{p\to\infty} ~~p D(m,m',p)= \lambda(m,m')~ ,
\ee
and its associated Poissonian GPF limit
\be
\Theta(z;m)= \lim_{p\to\infty} \Theta(z;m,p)~.
\ee 
Expression (\ref{nejjs})
leads to the nonlinear integral equation
\be
\Theta(z;m)= 
\displaystyle \exp\left[ \int_{m_0}^\infty dm'~
\lambda(m,m')(z \Theta(z;m')-1)\right]~ . 
\label{1}
\ee

It will be useful in the sequel to study the statistics of those
triggered events with magnitude larger than some threshold $\mu$, whose
corresponding  GPF for their numbers is denoted 
$\Theta(z;m,\mu)$. The equation for $\Theta(z;m,\mu)$ is obtained 
from (\ref{1}) by replacing $\Theta(z;m)$ by $\Theta(z;m,\mu)$ in both
the l.h.s. and r.h.s. and by replacing $z$ in the r.h.s. by
\be
H(\mu-m')+ z H(m'-\mu)=
\begin{cases}
1 & \text{if} \quad m'<\mu \\
z & \text{if} \quad m'> \mu~,
\end{cases}
\label{2}
\ee
where $H(x)$ is unit step function, equal to $1$ if $x>0$
and $0$ otherwise. Replacing $z$ by (\ref{2}) just
means that only those first generation
aftershocks, whose magnitudes are larger than $\mu$, are counted.
Their subsequent cascade above the 
magnitude level $\mu$ is accounted for by 
replacing $\Theta(z;m')$ by $\Theta(z;m',\mu)$. This leads to
\be
\begin{array}{c}
\Theta(z;m,\mu)=  \displaystyle \exp\big[
\int_{m_0}^\infty dm' \lambda(m,m')(
\Theta(z;m',\mu)-1)+ \\[3mm]
\displaystyle (z-1) \int_\mu^\infty dm' \lambda(m,m')
\Theta(z;m',\mu)\big]~ .
\end{array}
\label{3}
\ee

\subsection{Distribution of events magnitudes \label{pdfmethodda}}

One of the simplest and most informative
statistical characteristics of branching processes
is the average number $\langle R\rangle(m,\mu)$
of events of magnitude above $\mu$ triggered by some
spontaneous source of magnitude $m$:
\be
\langle R\rangle(m,\mu)= {d \Theta(z;m,\mu) \over dz}
\Big|_{z=1}~ .
\label{mgjls}
\ee
Using expression (\ref{3}) in (\ref{mgjls}), we find that $\langle
R\rangle(m,\mu)$ satisfies to linear integral equation
\be
\langle R\rangle(m,\mu)= \int_{m_0}^\infty \lambda(m,m')
\langle R\rangle(m',\mu)  dm'+ \langle R_1\rangle(m,\mu)~,
\label{4}
\ee
where
\be
\langle R_1\rangle(m,\mu)= \int_\mu^\infty \lambda(m,m') dm'
\ee
is the average number of first generation aftershocks with
magnitudes larger than $\mu$.

We assume that the spontaneous sources constitute a
stationary point process with Poisson statistics, with an average number of
spontaneous sources during a time interval $\tau$
equal to $\omega\tau$. Let us furthermore denote $p(m)$ the 
probability density function (PDF) of the magnitude of the 
random sources. Then, the total average number of events 
(including the spontaneous sources and all
their offsprings over all generations) during the time interval $\tau$
with magnitudes larger than $\mu$ is equal to
\be
\omega \tau [\langle R\rangle(\mu) +Q(\mu)] ~,
\label{5}
\ee
where
\be
\langle R\rangle(\mu)= \int_{m_s}^\infty \langle
R\rangle(m, \mu) p(m) dm 
\label{6}
\ee
is the average number of events of all generations with
magnitudes larger than or equal to $\mu$ which are triggered
by the spontaneous sources of all possible magnitudes
above some lower threshold $m_s$ defined as the smallest magnitude of spontaneous
events. In what follows, we assume that the 
spontaneous sources have their magnitudes distributed 
according to a GR law
\be
p(m)= \chi e^{-\chi (m-m_s)}~ H(m-m_s) ~ , 
\label{7}
\ee
with an exponent $\chi$ possibly distinct from those of triggered events.
The complementary cumulative distribution function (CDF) of spontaneous
sources magnitudes then reads
\be
Q(\mu)= \int_\mu^\infty p(m) dm= e^{-\chi (\mu-m_s)}~ ,
\qquad \mu>m_s~ . 
\label{8}
\ee

It is natural to introduce a magnitude threshold $m_d$ of
catalog completeness, i.e., only events with $m>m_d$ are observed.
Then, the total fraction of events above magnitude $\mu$
among all observable events in the time window $\tau$ is 
given by the following normalization of (\ref{5}) 
\be
F(\mu,m_d)= {\langle R\rangle(\mu) +Q(\mu) \over \langle
R\rangle(m_d) +Q(m_d)}~ .
\ee
$F(\mu,m_d)$ can be interpreted as the
complementary CDF of magnitudes of observable events. The
corresponding PDF of observable events is then
\be
f(\mu,m_d)= {g(\mu) +p(\mu) \over \langle R\rangle(m_d)
+Q(m_d)}~ , \qquad g(\mu)= - {d \langle R\rangle(\mu)
\over d \mu} ~. 
\label{9}
\ee
Expression (\ref{5}) also allows us to obtain the fraction 
of triggered events
\be
n(\mu)= { \langle R\rangle(\mu) \over \langle
R\rangle(\mu)+ Q(\mu)}~,
\label{10}
\ee
which is nothing but the average branching ratio \cite{HSnnn}.

In the sequel, we apply relations (\ref{9}) and (\ref{10}) to the
standard ETAS model (\ref{nhgjle}) and to the self-similar Vere-Jones model
(\ref{14}).

\section{Criticality condition and phase diagram}

\subsection{General maximum eigenvalue condition}

Before going further, it is important to derive the
conditions under which the branching process is not explosive,
i.e., for which the stationary spontaneous
sources point process generates a stationary sequence of triggered events.
The condition derives in general from an eigenvalue problem
(see \cite{Zhuangthesis} and references therein). 
In the present case, it is known that equation (\ref{4}) gives
bounded stationary solutions if the largest eigenvalue
$\rho$ of the corresponding homogeneous equation
\be
\rho \mathcal{R}(m)=\int_{m_0}^\infty \lambda(m,m')
\mathcal{R}(m') dm' 
\label{11}
\ee
is smaller than $1$ ($\rho<1$). This corresponds to the subcritical regime. The
condition $\rho=1$ defines the critical regime and $\rho>1$ gives
the explosive super-critical regime. In our analysis of the 
eigenvalue problem of equation (\ref{11}), we shall restrict to 
physically meaningful eigenfunctions $\mathcal{R}(m)$ 
which are monotonically increasing functions
growing no faster than the productivity law $\sim e^{a m}$:
\be
\lim_{m\to\infty} \mathcal{R}(m) e^{-a m} < \infty~ .
\label{12}
\ee

An important point should be noted in relation with Vere-Jones's model.
While in the standard ETAS model, the total number of triggered events
above any arbitrary magnitude $m_d$ diverges at criticality, we will see
that the average total number $\langle R\rangle(m_d)$ of observable
events is finite in the critical regime of Vere-Jones' model, while the
average total number $\langle R\rangle(m_d \to -\infty)$ is itself
infinite. Thus, while the process is critical when considering all
events of any magnitude, it becomes subcritical for events above a
finite threshold $m_d$. The hallmark of such subcritical regime is that
the fraction (\ref{10}) for observable events is smaller than $1$, i.e.
$n(\mu)<1$ for all $\mu>m_d$.

\subsection{Criticality condition for the standard ETAS model}

For the standard ETAS model defined in section \ref{etasstan}, equation (\ref{11}) reduces to
\be
\rho \mathcal{R}(m)=\kappa' e^{a (m-m_0)}
\int_{m_0}^\infty e^{-\beta (m'-m_0)} \mathcal{R}(m') d m'~ . 
\label{15}
\ee
Let us look for a solution of this equation in the form
\be
\mathcal{R}(m)=C\, e^{\delta m}~ .  \label{16}
\ee
Substituting (\ref{16}) in (\ref{15}) shows that $C \neq 0$ if and only if
$\delta=a$ while the corresponding eigenvalue is 
\be
\rho= \kappa \int_{m_0}^\infty e^{(a-\beta) m} dm=
{\kappa' \beta \over \beta-a}~ ,   \label{17} 
\ee
which is nothing but the average branching ratio $n$ defined in 
(\ref{mgmlwlws}). This recovers the known fact \cite{HS02} that the
condition $\rho=n<1$ corresponds to the sub-critical regime
associated with the solution 
\be
\langle R\rangle(m,\mu)=  {\kappa' \over 1- \rho}~ e^{a
(m-m_0)- \beta (\mu-m_0)}~ ,  \label{18}
\ee
of the non-homogeneous equation (\ref{4}).

Substituting (\ref{18}) and (\ref{7}) into (\ref{6}) and (\ref{9}), and 
assuming for simplicity that $m_s=m_0$, we obtain the
distribution of the PDF of
the magnitudes $\mu$ of observable events 
\be
f(\mu,m_d)= {\langle R\rangle \beta e^{-\beta (\mu-m_0)}+
\chi e^{-\chi (\mu-m_0)} \over \langle R\rangle e^{-\beta
(m_d-m_0)}+ e^{-\chi (m_d-m_0)}} H(\mu-m_d)~ , \label{19}
\ee
where
\be
\langle R\rangle= \langle R\rangle(m_0)={\rho' \over
1-\rho}~, \qquad \rho'=  {\kappa' \chi \over \chi-a}
\ee
is the average of the total number of aftershocks triggered
by one spontaneous source of arbitrary magnitude.
In the particular case where the GR laws for the magnitudes
of the spontaneous sources and of the
first generation aftershocks are the same,
i.e. if $\chi=\beta$, then the PDF of the magnitudes of observable events
given by (\ref{19}) reduces to the pure GR law
$f(\mu, m_d)= \beta e^{-\beta (m-m_d)} H(\mu-m_d)$.

Substituting
\be
\langle R\rangle(\mu)= {\rho' \over 1-\rho} e^{-\beta
(\mu-m_0)}~ , \qquad Q(\mu)= e^{-\chi(\mu-m_0)}
\ee
into expression (\ref{10}) yields
\be
n(\mu)= {\rho' \over \rho'+ (1-\rho) e^{(\beta-\chi)
(\mu-m_0)}}~ .
\ee
If, as before, $\chi= \beta$ ($\rho'=\rho$), then $n(\mu)$
does not depend on $\mu$ and is equal to the
average branching ratio $n=\rho$.

\subsection{Criticality condition for the self-similar Vere-Jones model}

For the self-similar Vere-Jones model defined in section \ref{jglld},
the homogeneous equation (\ref{11}) reduces to
\be
\rho \mathcal{R}(m)= \int_{-\infty}^\infty
\lambda(m-m') \mathcal{R}(m') d m' ~,
\ee
where $\lambda(m)$ is given by expression (\ref{14}). We search again a
solution for $\mathcal{R}(m)$ of the form (\ref{16}), which yields the
following eigenvalue $\rho(\delta)$ as a function of $\delta$ (shown
in figure 1):
\ba
\rho(\delta) &=& \int_{-\infty}^\infty \lambda(m) e^{-\delta m} dm=
{2\kappa d \over d^2- (\delta-\beta)^2}    \nonumber \\
&=& {s \over 1- (\delta -\beta)^2/d^2}~,  \label{20}
\ea
where
\be
s= {2 \kappa \over d}  \label{21}
\ee
is going to play an important and physically intuitive role in the sequel.
Note the major novelty compared with the standard ETAS model:
here, we obtain a continuous spectrum of eigenvalues $\rho(\delta)$
rather that the unique one (\ref{17}) associated with $\delta=a$
for the ETAS model.  

In expression (\ref{20}), since $1- (\delta -\beta)^2/d^2 \leq 1$, 
there is a solution with
$\rho(\delta) \leq 1$ associated with the sub-critical
and critical regimes only for $s \leq 1$. For $s>1$, all eigenvalues
$\rho(\delta)$ are larger than $1$, which corresponds to the
explosive super-critical regime.  We show below
that the parameter $s$ plays the role of an average branching ratio.

For $s \leq 1$, the continuous spectrum of
eigenvalues is indexed by $\delta$ spanning
the interval $[\beta -h, \beta]$, for which $\rho(\delta) \leq 1$
characterizes the sub-critical and critical regimes. The exponent $h$
is determined by the condition $\rho(\beta-h)=1$, whose 
geometrical determination is represented in figure 1. We rule out
the possibility $\delta > \beta$, which leads to unphysical 
solutions as shown below. Given this spectrum of eigenvalues,
the growth of $\mathcal{R}(m)$ given by (\ref{16}) is controlled
by the largest eigenvalue $\rho(\beta - h)=1$, when it exists,
which leads to
\be
\mathcal{R}(m)\sim e^{(\beta- h) m}~, \qquad h=d \sqrt{1-s} ~ . \label{22}
\ee

We can now describe the phase diagram of the self-similar
Vere-Jones model, shown in figure 2.
\begin{itemize}
\item For $s>1$, all eigenvalues $\rho(\delta)$ are larger than $1$,
corresponding to the explosive super-critical regime.

\item For $s<1$ and $d<\beta$, we accept (\ref{22}) as a physically appropriate
eigenfunction only if it is monotonically increasing with respect to
$m$, that is, if $0 \leq h \leq \beta$. This leads to the condition 
\be
 {1 \over 2} \left(d- {\beta^2 \over d}\right) <\kappa < {d \over 2}~.
 \label{24}
\ee
In the range (\ref{24}) of parameters, there
is always a solution, whatever
the value $0 \leq s \leq 1$ of the form
(\ref{22}), associated with the unit eigenvalue. 
This is the critical regime, which is associated with a
finite range of parameters $0 \leq s \leq 1$ and $0 \leq h \leq \beta$
corresponding to the domain (\ref{24}), indicated in figure 2.

\item For 
\be
0< \kappa < {1 \over 2} \left(d- {\beta^2 \over d}\right) ~,
\label{23}
\ee
which is only possible if $d>\beta$,
we obtain $h>\beta$, that is, $\delta < 0$, which corresponds
to a seismic activity which is a decreasing 
function of the mother magnitude. This is the
hallmark of the subcritical regime. We will see below that
the average of the total number of offsprings is finite in this
sub-critical regime.
\end{itemize}
Figure 2 summarizes this phase diagram and delineates
the domains of existence of the 
sub-critical, critical and super-critical regimes in the
plane of parameters ($\kappa,d$) for a given $\beta$.

The physical meaning of this classification is obtained by
examining the equation for the average number $\langle R\rangle(m,\mu)$
of aftershocks which are triggered by some mainshock with
magnitude $m$. This equation is expression (\ref{4}) written
for the self-similar Vere-Jones model:
\be
\langle R\rangle(m,\mu)= \int_{-\infty}^\infty
\lambda(m-m') \langle R\rangle(m',\mu)  d m'+ \langle
R_1\rangle(m,\mu) ~,  \label{25} 
\ee
where
\be
\langle R_1\rangle(m,\mu)= \int_\mu^\infty \lambda(m-m')
d m'~    \label{26}
\ee
is the average of the number of corresponding first-generation
aftershocks.
Let us introduce the auxiliary function
\be
S(m,\mu)= - {\partial \langle R\rangle(m,\mu) \over \partial \mu}~.
\ee
Using (\ref{25}), $S(m,\mu)$ is solution of
\be
S(m,\mu)= \int_{-\infty}^\infty \lambda(m-m') S(m',\mu) d
m'+ \lambda(m-\mu)~ .
\ee
For its structure, it is clear that 
the solution of this equation depends only on
the difference between $m$ and $\mu$, so that we write
\be
S(m,\mu)= S(m-\mu)~. 
\ee
The value of $\langle R\rangle(m,\mu)$ is obtained from $S(m,\mu)$ by using 
\be
\langle R\rangle(m,\mu)\equiv \int_\mu^\infty S(m,\mu')
d\mu'= \int_{-\infty}^{m-\mu} S(x) dx~ .  \label{27}
\ee

Let us solve the equation
\be
S(m)= \int_{-\infty}^\infty \lambda(m-m') S(m') d m'+ \lambda(m)
\ee
for the auxiliary function $S(m)$. Applying 
the two-sided Laplace transform to this equation yields
an equation for the Laplace transform
\be
\hat{S}(u)= \int_{-\infty}^\infty S(m) e^{-u m} dm
\ee
which takes the following form:
\be
\hat{S}(u)= {\hat{\lambda}(u)   \over 1- \hat{\lambda}(u)}~ . 
\label{28}
\ee
The Laplace transform $\hat{\lambda}(u)$ of the kernel
$\lambda(m)$ given by (\ref{14}) can be explicitly calculated as
\be
\hat{\lambda}(u)= {2\kappa d \over d^2- (u-\beta)^2}~ .
\label{29}
\ee
Substituting expression (\ref{29}) into (\ref{28}) yields
\be
\hat{S}(u)= {2\kappa d \over h^2- (u-\beta)^2}~ , \qquad
h^2= d^2- 2\kappa d~ .
\ee
Its inverse Laplace transform
\be
S(m)= {1 \over 2\pi i} \int_{\beta- i\infty}^{\beta+ i
\infty} \hat{S}(u) e^{u m} d u
\ee
gives finally
\be
S(m)= {\kappa d \over h}~ e^{\beta m- h |m|}~ . 
\label{30}
\ee

We can now use (\ref{30}) in (\ref{27}) to characterize the
average of the number of aftershocks in the sub-critical (\ref{23}), critical
(\ref{24}) and super-critical ($s>1$) regimes.
\begin{itemize}
\item In the subcritical case $h>\beta$, we can take 
$\mu \to -\infty$ and still obtain a finite limit 
\be
\langle R\rangle= \lim_{\mu\to-\infty} \langle
R\rangle(m,\mu)= \int_{-\infty}^\infty S(x) dx= {2 \kappa
d \over h^2-\beta^2}< \infty 
\label{31}
\ee
for the average of the total number of aftershocks
of all generations triggered by an
arbitrary spontaneous source. Thus, in the subcritical regime, the averages of
total number of both observable and unobservable events are finite.

\item In the critical regime (\ref{24}), we find that the average
$\langle R\rangle(m,\mu)$ of the total number of events above any finite
magnitude threshold $\mu$ is finite. In contrast, the average of the
total number of aftershocks (including the unobservable tiny events of
magnitudes $m' \to -\infty$) becomes infinite: $\langle R\rangle=\infty$. 
It is remarkable that we have at the same time $\langle R\rangle(m,\mu)$ finite
for any $\mu > -\infty$ and $\langle R\rangle=\infty$. In real data, 
we only observe $\langle R\rangle(m,\mu)$. The underlying criticality
is thus unobservable: due to the special cancellations in the self-similar
Vere-Jones model, the infinite swarm of tiny events form an
unobservable sea of activity, whose observable consequences lie
in the finite activity at finite magnitudes. Thus, in a sense,
this critical regime (\ref{24}) can actually be decomposed into
an effective sub-critical regime for $s<1$ and a critical point reached at $s=1$
for observable events. Indeed, as 
$s\to 1$ (i.e. if $h\to 0$), expressions
(\ref{27}) and (\ref{30}) for the average number of observable aftershocks
triggered by an arbitrary spontaneous source tends to
infinity. This shows that the critical regime for
observable events corresponds to $s=1$ and confirms the interpretation
of $s$ as the effective branching ratio for observable events.

\item As $s\to 1$ ($h\to 0$), $\langle
R\rangle(m,\mu)$ tends to infinity for any $m$ and $\mu$,
confirming that the super-critical regime corresponds to $s>1$.
\end{itemize}

\subsection{Criticality condition for the generalized Vere-Jones ETAS model}

We now analyze the critical conditions for the generalized Vere-Jones 
ETAS model defined by expression (\ref{13}) in section \ref{genevj}.
It is convenient for the analysis to represent
the intensity $\lambda(m,m')$ in (\ref{13}) by
\be
\lambda(m,m')= e^{a(m-m')} \nu(m,m')~ ,   \label{mlss}
\ee
where
\be
\nu(m,m')=
\kappa e^{(a-\beta) m'- d |m-m'|}~.    \label{mlss2}
\ee
Let us introduce the auxiliary function
\be
\mathcal{U}(m)= \mathcal{R}(m) e^{-a m} ~,
\label{44}
\ee
and rewrite the homogeneous equation (\ref{11}) in the form
\be
\rho \mathcal{U}(m)=\int_{m_0}^\infty \mathcal{U}(m')
\nu(m,m') dm' ~ .  \label{45}
\ee
From the definition of $\nu(m,m')$ in (\ref{mlss2}), we see that
the following equality holds:
\be
{d^2 \nu(m,m') \over dm^2}=d^2 \nu(m,m')- 2 \kappa d~
e^{(a-\beta)m}~ \delta(m'-m)~ ,   \label{46}
\ee
where $\delta(x)$ is the Dirac-delta function.
By differentiating equation (\ref{45}) twice with respect to
$m$, the identity (\ref{46}) leads to the condition that, if
$\mathcal{U}(m)$ is a solution of Eq.~(\ref{45}), then
it should also satisfy the differential equation
\be
\rho {d^2 \mathcal{U}(m) \over dm^2}=\left[\rho d^2- 2
\kappa d e^{(a-\beta)m} \right] \mathcal{U}(m)~ , \qquad
m>m_0~ .   \label{47}
\ee
Thus, to determine the eigenvalue of equation (\ref{45}), our 
strategy is to search for a solution of equation (\ref{47}),
which at the same time satisfies the integral condition
\be
\rho \mathcal{U}(m_0)= \kappa \int_{m_0}^\infty
\mathcal{U}(m) e^{(a-\beta) m- d(m-m_0)} dm~.  \label{48}
\ee
which is derived from (\ref{45}). It is straightforward
to check that this strategy leads to solving (\ref{45}).
In addition, we need to impose that the eigenfunction $\mathcal{R}(m)$
is monotonically increasing as a function of $m$ such that the 
condition (\ref{12}) is satisfied. In terms of the auxiliary
function $\mathcal{U}(m)$, this implies that
\be
\lim_{m\to\infty} \mathcal{U}(m)< \infty ~ .   \label{49}
\ee
Thus, our problem is to find a solution of (\ref{47})
with the two conditions (\ref{48}) and (\ref{49}).

Before using this formalism for the generalized Vere-Jones 
ETAS model, it is instructive to see how it performs on the
standard ETAS model (corresponding to $d=0$). For $d=0$,
equation (\ref{47}) reduces to
\be
\rho {d^2 \mathcal{U}(m) \over dm^2}= 0~ .
\ee
Its solution satisfying condition (\ref{49}) is
$\mathcal{U}= C=\text{const}$. Substituting it into the
integral relation (\ref{48}) recovers the known
expression for the unique eigenvalue equal to the 
average branching ratio:
\be
\rho= {\kappa e^{(a-\beta)m_0} \over \beta-a}= {\gamma
\over \gamma-1}~\kappa'~, \qquad \gamma= {\beta \over a} ~ .
\ee
The critical regime corresponds to the set of parameters obeying
\be
\kappa'= \mathcal{K}(a, \beta)= 1- {a \over \beta}~,
\label{50}
\ee
while the domain of subcritical regime corresponds to $\kappa <
\mathcal{K}(a, \beta)$.
Similarly, to explore the condition for criticality in the generalized Vere-Jones
ETAS model, we just need to put $\rho=1$ in (\ref{47})
and search for the function 
\be
\kappa= \mathcal{K}(a,\beta,d)
\ee
such that the homogeneous equation
\be
{d^2 \mathcal{U}(m) \over dm^2}=\left[d^2- 2 \kappa d
e^{(a-\beta)m} \right] \mathcal{U}(m)~ , \qquad m>m_0~ ,
\label{51}
\ee
has a nontrivial solution increasing with $m$, which satisfies
the integral equality (\ref{48}) expressed for $\rho=1$.
The sub-critical regime then corresponds to the set of parameters
such that $\kappa< \mathcal{K}(a, \beta, d)$.

In order to solve (\ref{51}) for the critical case $\rho=1$, 
we introduce the new function $v(y)$ such that
\be
\mathcal{U}(m)= v(y)~, \label{52}
\ee
with
\be
y=\varrho e^{-(m-m_0) d/\epsilon}~, \qquad\varrho=
{\sqrt{8\kappa' d \beta} \over \beta-a} ~ , ~~~
\epsilon= {2 d \over \beta -a} ~ .   \label{53}
\ee
Note that the parameter $\epsilon$ quantifies the transition
from the ETAS model (obtained for $\epsilon=0$) to the self-similar
Vere-Jones model (obtained for $\epsilon \to +\infty$).
For small (resp. large) $\epsilon$, the generalized Vere-Jones ETAS model
is close to the standard ETAS (resp. self-similar Vere-Jones) model.
Equation (\ref{51}) for $\mathcal{U}(m)$ translates into the following 
Bessel equation for $v(y)$: 
\be
y^2 {d^2 v(y) \over dy^2}+y { d v(y) \over dy} +(y^2-
\epsilon^2) v(y)=0 ~ .   \label{54}
\ee
It follows from (\ref{49}) and from the definition (\ref{53}) of $y$ that 
the solution of (\ref{54}) has to satisfy
\be
\lim_{y\to 0} v(y)<\infty~ .
\ee
The integral condition (\ref{48}) imposes in addition that
\be
v(\varrho)= {1 \over 2}~ \epsilon^{-1} ~ \varrho^{-
\epsilon} \int_0^{\varrho} v(y) y^{\epsilon +1} dy~ .
\label{55}
\ee
The solution of  Eq.~(\ref{54}) has the form of a Bessel function
\be
v(y)= A J_\epsilon (y)~ ,   \label{56}
\ee
where $A$ is a constant.
Substituting (\ref{56}) into (\ref{55}) and using the known
recursion relation
\be
\int y^{\epsilon+ 1} J_\epsilon(y) ~ dy = y^{\epsilon+ 1}
J_{\epsilon+1}(y)
\ee
between Bessel functions, we obtain the implicit equation 
for the variable $\varrho$
\be
J_\epsilon(\varrho)= {\varrho \over 2 \epsilon}
J_{\epsilon+1}(\varrho)~ , \label{57}
\ee
which determines the set of parameters corresponding to criticality ($\rho=1$).

Further insight can be obtained by determining the 
leading behavior of $\varrho$ for small $\epsilon$ (quasi-ETAS model)
and large $\epsilon$ (quasi-self-similar Vere-Jones model). For this,
we introduce the new auxiliary parameter
\be
\psi= {\varrho \over \epsilon}= \sqrt{{2 \kappa' \beta
\over d}} ~,  \label{58}
\ee
and rewrite equation (\ref{57}) in the form
\be
\psi = {2 J_\epsilon(\epsilon \psi) \over J_{\epsilon+1}
(\epsilon \psi)} ~ .  \label{59}
\ee
For $\epsilon\ll 1$, we use the expansion
\be
{2 J_\epsilon(x) \over J_{\epsilon+1} (x)} \simeq {4
\over x} (1+\epsilon) - {x \over 2+\epsilon} \qquad (x\ll 1)
\ee
which gives the solution
\be
\psi \simeq \sqrt{{2  \over \epsilon}~ (2+\epsilon)}
\qquad (\epsilon\ll 1) ~.   \label{60}
\ee
In term of the original parameters $\kappa$, $a$ and $d$, this gives
\be
\kappa' \simeq 1-{a \over \beta}+ {d  \over \beta}~ 
\label{61}
\ee
as the relation expression the critical regime $\rho=1$ of the 
generalized Vere-Jones ETAS model. Expression (\ref{61}) differs
from its counterpart (\ref{50}) obtained for the standard ETAS model
by the correction term $d/\beta$, which describes 
a kind of broadening of the subcritical
regime due to the magnitude localization effect ($d>0$).

In the other limit $\epsilon\gg 1$ corresponding to the quasi-Vere-Jones
model, the solution of the criticality condition (\ref{59}) can also be
obtained asymptotically by expanding the Bessel functions 
in the neighborhood of their first zero, i.e., by searching for 
$\epsilon \rho$ close to $\upsilon=\upsilon(\epsilon)$ defined by
$J_\epsilon[\upsilon(\epsilon)]=0$. The corresponding
expansions in Taylor series of the Bessel functions are
\be
J_\epsilon(x) \simeq J'_\epsilon(\upsilon) (x-\upsilon)~,
~~ J_{\epsilon+1}(x) \simeq J_{\epsilon+1}(\upsilon)+
J'_{\epsilon+1}(\upsilon) (x-\upsilon)~ .  \label{63}
\ee
It is well-known from the theory of Bessel functions that
\be
J'_\epsilon(\upsilon)= -J_{\epsilon+1}(\upsilon) ~ ,
\qquad J'_{\epsilon+1}(\upsilon)= -{\epsilon+1 \over
\upsilon}J_{\epsilon+1}(\upsilon) ~ .   \label{64}
\ee
Thus, for $\epsilon\gg 1$, expressions (\ref{63}) and (\ref{64}) allow us
to replace the criticality condition (\ref{59}) by
its asymptotic expression
\be
\psi \simeq {2 \upsilon ( \upsilon- \epsilon \psi ) \over
\upsilon+ (\epsilon+1) ( \upsilon- \epsilon \psi)}~.
\ee
Solving this equation in $\psi$ yields
\be
\psi \simeq {\upsilon(\epsilon) \over 1+\epsilon} \qquad
(\epsilon\gg 1) ~.  \label{65}
\ee
Using the known asymptotic relation 
\be
\upsilon(\epsilon)\simeq \epsilon+1.856~ \epsilon^{1/3}
+1.033 \epsilon^{-1/3}    \label{66}
\ee
for the first-zero of the Bessel function of order $\epsilon$,
the criticality condition formulated in terms of the
original parameters $\kappa$, $a$, $d$ then reads
\be
\kappa' \simeq {d \over 2 \beta}+ {0.58 \over \beta}
d^{1/3}~ (\beta-a)^{2/3}~ .    \label{67}
\ee
The first leading term $\kappa' \approx {d \over 2 \beta}$ 
recovers the critical condition $s=2\kappa/d=1$ for the 
self-similar Vere-Jones model. The last term in the r.h.s.
of (\ref{67}) thus provides the first correction to the 
exactly self-similar Vere-Jones model when $a \neq \beta$.
Figure 3 plots the numerical solution of equation (\ref{59}) 
as a function of $\epsilon$ together with its two 
asymptotics regimes (\ref{60}) and (\ref{65}) with (\ref{66}).

\section{Distribution of earthquake magnitudes for Vere-Jones's self-similar model}

The Gutenberg-Richter distribution of earthquake magnitudes
is perhaps the most ubiquitous and documented statistical
property of earthquake catalogs. It is thus of great
interest to investigate the prediction of Vere-Jones self-similar
model of this quantity. We analyze in turn two versions of this statistics.
first when the magnitude of the 
mainshock is known and second when summing over all possible magnitudes
as in a large catalog. Both forms of the distributions have been 
investigated in the empirical literature. 

\subsection{Distribution of the magnitude of aftershocks triggered
by a mainshock of given magnitude $m$}

Reasoning as in section \ref{pdfmethodda} shows that the PDF's of 
the magnitudes $\mu$ of first-generation aftershocks and 
of all aftershocks of all generations 
of a mainshock of magnitude $m$ are proportional
respectively to $S_1(m,\mu)$ and $S(m,\mu)$ given by
\be
S_1(m,\mu)= -{\partial \langle R_1\rangle(m,\mu) \over
\partial \mu}~ , ~
S(m,\mu)= -{\partial \langle R\rangle(m,\mu) \over
\partial \mu}~ .
\ee
Their expressions are
\ba
S_1(m,\mu) &=& \kappa e^{-\beta (\mu-m)- d|\mu-m|}~ , \\
S(m,\mu) &=& {\kappa d \over h} e^{-\beta (\mu-m)- h|\mu-m|}~ . \label{68}
\ea
Thus, the two branches $\mu<m$ with exponent $\beta -d$ and $\mu>m$ with
exponent $\beta + d$ of the PDF $S_1(m,\mu)$ of first-generation
aftershocks are renormalized in two other branches of the same form,
with $d$ renormalized into $h=d \sqrt{1-s}$. The main point is that only
$S(m,\mu)$ is observed in real catalogs since one does not have the luxury
of being able to distinguish between the different generations of
triggered aftershocks. Expression (\ref{68}) shows that the PDF $S(m,\mu)$
may be quite close to a single power law, even if the difference between
the two branches of $S_1(m,\mu)$ is significant, as long as the critical
parameter $s$ is close to $1$ (for $s\to 1$, 
$S(m,\mu)\sim e^{-\beta \mu}$ and is a pure power law). In other words, an observable PDF of
aftershock magnitudes close to a single power law is compatible with
strong deviations from a pure power law for first-generation events, due
to the renormalization effect over all the generations which effectively
mixes up the two branches sufficiently close to criticality $s \to 1$.
Thus, as in previous studies of the standard ETAS model
\cite{SSBaththe,SaiSorpdf}, comparison between the data and these
predictions suggests that the earth is operating close to a critical point.

\subsection{Distribution of earthquake magnitudes over all events}

The Gutenberg-Richter law for the distribution of the magnitudes of
earthquakes is generally a statistical property established
for a large space-time-magnitude domain, without restrictions. 
It is thus interesting to ask what Vere-Jones' self-similar model
predicts for the distribution of magnitude of a large stationary sequence
of events triggered by a steady-state influx of spontaneous sources.
The answer is given by expression (\ref{9}) for 
$f(\mu,m_d)$. We thus need to make explicit the dependence of 
$f(\mu,m_d)$ on the parameters of the model. This problem depends on
four key parameters: $\chi$ is the exponent 
for spontaneous sources defined in (\ref{7}); $\beta$ is a crucial
parameter in the definition of the PDF of magnitudes of first-generation
aftershocks; $h$ and $s$ appear in the condition for criticality.

Let us first investigate the contribution $g(\mu)$
of the events triggered by the 
spontaneous source. For Vere-Jones' self-similar model, we have 
\be
g(\mu)= \int_{m_s}^\infty S(m-\mu) p(m) dm~ .  \label{mgmel}
\ee
Substituting (\ref{7}) and (\ref{30}) in (\ref{mgmel}) yields
\be
g(\mu)= p(\mu) K(m_s-\mu, \beta-\chi)~,  \label{32}
\ee
where $K(m_s-\mu, \beta-\chi)$ describes the deviation of the PDF
of the magnitudes of triggered events from the PDF of the magnitudes
of the spontaneous sources. This function is given by
\be
K(y,z)= {\kappa d \over h} \int_y^\infty e^{ z x-h|x|}
dx~ .   \label{33}
\ee
Its explicit expression is
\be
K(y,z)= {2\kappa d \over h^2- z^2} - {\kappa d \over
h(h+z)} e^{(z+h) y}~ . \label{38}
\ee
The corresponding complementary cumulative distribution, i.e., 
the total number of triggered events with magnitude larger than 
$\mu$, is equal to
\be
\langle R\rangle(\mu)= \int_{\mu}^\infty g(m) dm=
\int_\mu^\infty p(m) K(m_s-m,\beta- \chi) dm~ . \label{34}
\ee

We need to distinguish three cases as $g(\mu)$ given by (\ref{32}) is qualitatively
different for different values $\chi, \beta$ and $h$.
\begin{itemize}
\item For $\chi<\beta- h$, $g(\mu)=\infty$ for any $\mu$. This results
from the fact that the spontaneous sources with large magnitudes
dominate the production of triggered events in this case. This super-critical regime
can be tamed with the introduction of an upper magnitude cut-off $m_{\rm
max}$ but is not investigated further here. The limiting case
$\chi=\beta-h$ gives a criticality condition for the observable events
in the framework of the self-similar Vere--Jones model.

\item Consider the regime
\be
\beta- h < \chi <\beta + h~,
\label{36}
\ee
and the limit $m_s \to -\infty$ corresponding to $y \to -\infty$, for which $K(y,z)$
has the following asymptotic dependence
\be
K(y,z)\simeq {2\kappa d \over h^2- z^2}~ , \qquad y\to -\infty~ .
\ee
The corresponding asymptotic expressions for the PDF $g(\mu)$ given
by (\ref{32}) and the complementary cumulative distribution
given by (\ref{34}) are
\be
g(\mu)\simeq {2\kappa d \over h^2- z^2} p(\mu)~ , ~
\langle R\rangle(\mu) \simeq {2\kappa d \over h^2- z^2}
Q(\mu)~.   \label{39}
\ee
Substituting (\ref{39}), (\ref{7}) and (\ref{8}) into (\ref{9})
yields the distribution of the magnitudes of all observable events
(in the limit $m_s\to -\infty$)
\be
f(\mu,m_d)= \chi e^{-\chi(m-m_d)} \qquad (\mu>m_d)~ .
\label{40}
\ee
Remarkably, in this regime (\ref{36}), the observed
Gutenberg-Richter distribution is predicted to reveal only
the exponent of the distribution of the spontaneous sources
and to be blind to the exponents $\beta, d$ of the distribution
of triggered events. In other words, the PDF of the magnitudes
of all observed events reproduces that of the spontaneous sources given by (\ref{7}).

Substituting (\ref{39}) into (\ref{10}) yields the
fraction of triggered events whose magnitudes are large than
$\mu$, among all analogous events,
\be
n(\mu)=n= {2\kappa d \over d^2- (\beta-\chi)^2}~ ,
\label{41}
\ee
which is found independent of $\mu$. If the same distribution of 
magnitudes describes the spontaneous sources and the triggered events
($\beta=\chi$), then $n=s$. Thus, the critical regime
corresponds to $s=1$, confirming the interpretation of $s$ as
equivalent to the critical branching ratio of Vere-Jones' self-similar model.
However, if $\beta \neq \chi$ and $|\beta-\chi|\to h$, then
we obtain $n\to 1$ even for $s<1$ ($h>0$): essentially all the events are
triggered.

\item Consider the regime
\be
\chi > \beta + h~.
\label{37}
\ee
In this case, the function $K$ is closely approximated by its asymptotic behavior
\be
K(y,z)\simeq -{\kappa d \over h(h+z)} e^{(z+h) y}~ .
\label{42}
\ee
The corresponding asymptotic behaviors of $g(\mu)$ given by (\ref{32}) and $\langle
R\rangle(\mu)$ given by (\ref{34}) for large $\mu-m_s$ are
\be
g(\mu) \simeq {\kappa d \chi \over h(\chi-\beta-h)}
e^{-(\beta+h)(\mu- m_s)}
\ee
and
\be
\langle R\rangle(\mu) \simeq {\kappa d \chi \over
h(\chi-\beta-h)(\beta+h)} e^{-(\beta+h)(\mu- m_s)}~.
\ee
Substituting these two expression into (\ref{9}) and (\ref{10}) yields
\be
f(\mu,m_d)\simeq (\beta+h) e^{-(\beta+h)(\mu-m_d)}~ ,
\qquad n\simeq 1~ ,  \label{43}
\ee
asymptotically for $\mu- m_s\to +\infty$.
\end{itemize}

\section{Summary and discussion}

Our results can be summarized as follow.

$\bullet$ We have clarified and quantified the conditions under which the
self-similar Vere-Jones model as well as a more general version (which
contains both the standard ETAS and Vere-Jones version as special cases)
are critical, subcritical and supercritical. Only the subcritical
and critical regimes give a stationary process in the presence
of a non-zero flux of immigrants.

$\bullet$ We have shown that the concept of an average branching ratio $s$, defined as
the average number of daughters of first generation per mother of
magnitudes above a finite magnitude threshold) holds for Vere-Jones model in a broad
domain of parameters. Remarkably,
$s$ is found independent of the magnitude threshold used.
However, the average branching ratio loses its meaning when the
magnitude threshold is pushed to $-\infty$
(in other words, when it is removed), as the existence of arbitrary small events
allowed in this model dominates and makes the average divergent. 
Since empirical catalogs are always characterized
by a minimum magnitude $m_d$ of completeness, our results apply directly
and show that it is possible to have an infinite number of -- unobservable
but still important for the cascade of triggering -- events per mother
together with a finite average branching ratio for observable events.

$\bullet$ Vere-Jones' model is defined by the Gutenberg-Richter (GR)
magnitude distribution for first-generation events triggered by a source
of magnitude $m$ having two branches: for aftershocks magnitudes $m'<m$,
the GR exponent is $\beta -d$, while it is $\beta + d$ for $m'>m$. We have
shown that, accounting for the contributions of all generations of triggered
events, this GR distribution is renormalized into another two-branches
law: for aftershocks magnitudes $m'<m$, the renormalized GR exponent is
$\beta -h$, while it is $\beta + h$ for $m'>m$, where $0 \leq h=d
\sqrt{1-s} \leq d$ for $0 \leq s \leq 1$. 

$\bullet$ Since only the renormalized GR 
is observable in the analysis of empirical catalogs, the fact that the literature
reports a single exponent $\beta$ for the GR distribution implies that, 
assuming that Vere-Jones model is a correct description, $h$ is small so that the
difference between the exponents $\beta-h$ and $\beta +h$ is within
the empirical uncertainties and variations from catalog to catalog. This implies
that either $d$ is small or $s$ is close to $1$ (condition
describing the boundary between the subcritical to supercritical regimes
of the Vere-Jones model) or both. We note that the prediction that two
distinct exponents $h$ and $d$ characterize respectively the GR law over
all generations and the GR law over the first generation of triggered events
could  be in principle tested empirically using an adaptation of the
statistical declustering method developed by Zhuang et al. \cite{Zhuang1,Zhuang2}
to Vere-Jones's model. In this respect, Zhuang et al. \cite{Zhuang2} have 
already found that the magnitude distribution
of the triggered event depends on the magnitude of its direct ancestor, with 
an exponent smaller for large events (in contradiction with 
other less sophisticated studies using more arbitrary space-time 
windows \cite{HelmPRL,HelmKaganJackson04}): this is roughly consistent with our 
prediction with Vere-Jones model, as the observable distribution for
large (resp. small) ancestors is weighted more by the $m'<m$ (resp. $m'>m$) regime associated
with exponent $\beta -h$ (resp. $\beta +h$).

$\bullet$ The distribution of magnitudes over a stationary catalog (obtained
by summing over an average steady flow of spontaneous sources) is found
universal (independent of $\beta$, $d$ and the other parameters) and a
pure GR with exponent equal to the exponent $\chi$ of the spontaneous
sources (which in full generality is allowed to be different from the
exponent $\beta$ involved in the distribution of triggered events) in 
a large domain of the parameter space. This
implies that, if the exponent of the spontaneous sources is different
from the exponent of triggered events, the physics of cascades of
triggering in the self-similar Vere-Jones model implies that only the
former exponent is observable in global catalogs. Again, the
statistical declustering method of Zhuang et al. \cite{Zhuang1,Zhuang2} should
be able to test this prediction. In this respect, we note that Zhuang et al.
find that the background events have a larger exponent than the triggered events:
$\chi > \beta$.

{\bf Acknowledgments:} This work is partially supported
by NSF-EAR02-30429, and by the Southern California
Earthquake Center (SCEC) SCEC is funded by NSF
Cooperative Agreement EAR-0106924 and USGS Cooperative
Agreement 02HQAG0008. The SCEC contribution number for
this paper is xxx.

{}

\clearpage

\begin{quote}
\centerline{
\includegraphics[width=14cm]{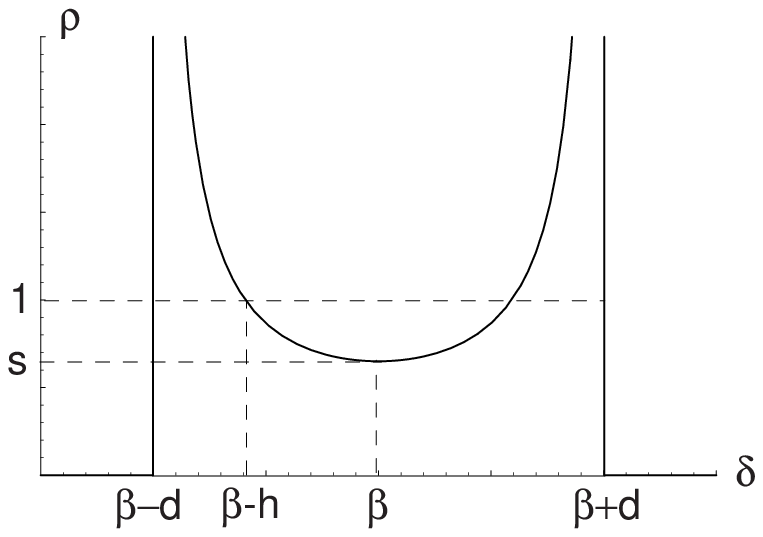}}
{\bf Fig.~1:} \small{Dependence of the eigenvalue $\rho$
given by (\ref{20}) associated with the eigenfunction $\mathcal{R}(m)$ (\ref{16}),
as a function of $\delta$ for the self-similar Vere-Jones model. }
\end{quote}

\clearpage

\begin{quote}
\centerline{
\includegraphics[width=14cm]{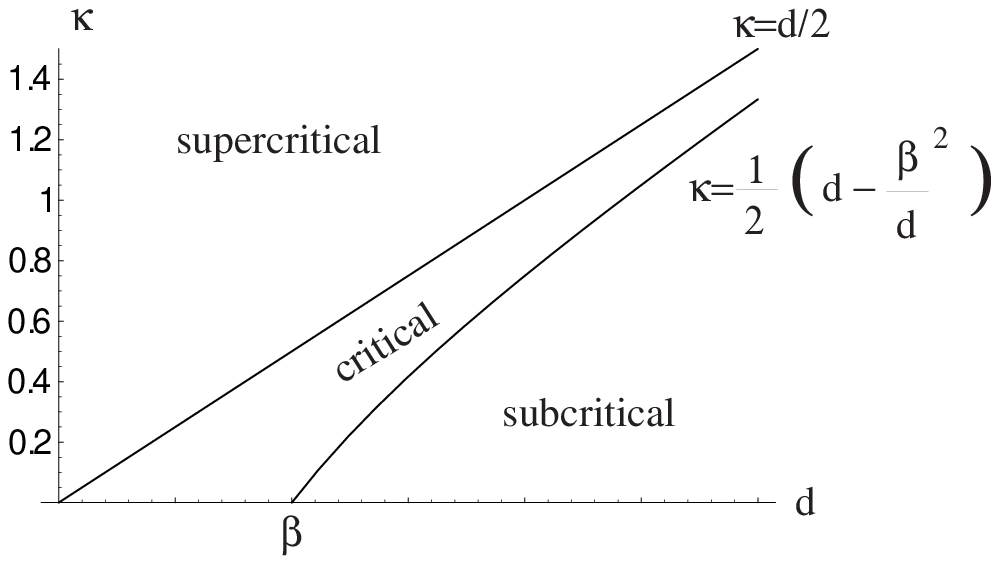}}
{\bf Fig.~2:} \small{Phase diagram in the plane of parameters ($\kappa,d$) 
at fixed $\beta$, for the self-similar Vere-Jones model, 
showing the domains of existence of the
sub-critical (\ref{23}), critical (\ref{24}) and super-critical ($s>1$)
regimes.}
\end{quote}

\clearpage

\begin{quote}
\centerline{
\includegraphics[width=14cm]{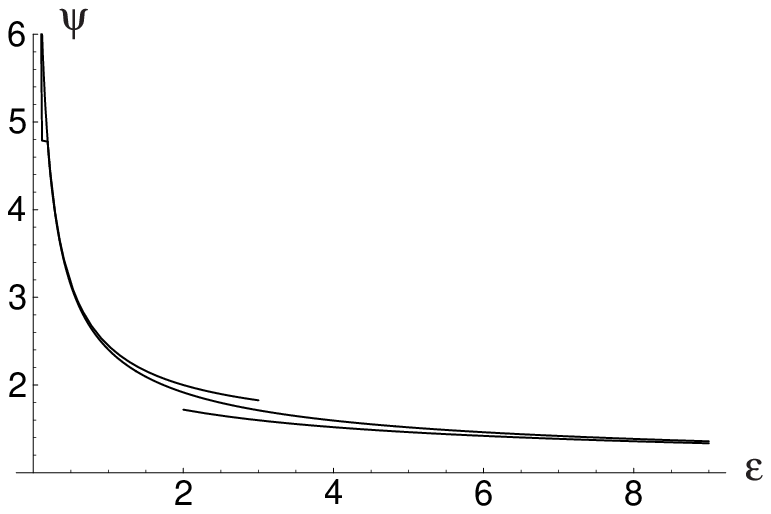}}
{\bf Fig.~3:} \small{Numerical solution of equation (\ref{59}) 
as a function of $\epsilon$ defined in (\ref{53}) compared with its two 
asymptotics regimes (\ref{60}) and (\ref{65}) with (\ref{66}),
giving the condition for criticality $\rho=1$ for 
the generalized Vere-Jones ETAS model.}
\end{quote}

\end{document}